\documentclass{Interspeech}

\interspeechcameraready

\title{On the Within-class Variation Issue in Alzheimer's Disease Detection}

\author[affiliation={1}]{Jiawen}{Kang}
\author[affiliation={2}]{Dongrui}{Han}
\author[affiliation={1}]{Lingwei}{Meng}
\author[affiliation={1}]{Jingyan}{Zhou}
\author[affiliation={1}]{Jinchao}{Li}
\author[affiliation={1}]{Xixin}{Wu}
\author[affiliation={1,2}]{\quad \quad \quad Helen}{Meng}

\affiliation{}{The Chinese University of Hong Kong}{Hong Kong SAR, China}
\affiliation{}{Centre for Perceptual and Interactive Intelligence}{Hong Kong SAR, China}

\email{jwkang@se.cuhk.edu.hk}


\usepackage{comment}
\usepackage{cite}
\usepackage{amsmath,amssymb,amsfonts}
\usepackage{algorithmic}
\usepackage{graphicx}
\usepackage{textcomp}
\usepackage{multirow}
\usepackage{booktabs}
\usepackage{stfloats}
\usepackage[dvipsnames]{xcolor}
\usepackage{array}

\usepackage{amsmath,amssymb,amsthm}
\theoremstyle{definition}   

\begin{document}

\maketitle

\begin{abstract}
Alzheimer's Disease (AD) detection commonly employs machine learning
classification models to distinguish between individuals with AD and
those without. Different from conventional classification tasks, AD
detection involves substantial within-class variation, as individuals
sharing the same diagnosis may exhibit different degrees of cognitive
impairment. We formulate two aspects of this issue: within-class
heterogeneity and instance-level imbalance. To model such variation
under binary supervision, we estimate sample-specific AD class
probabilities as sample scores and develop two corresponding methods:
Soft Target Distillation (SoTD) and Instance-level Re-balancing
(InRe). Experiments on the ADReSS and CU-MARVEL corpora show that the
estimated scores align with independent cognitive assessments and that
the proposed approaches improve AD detection performance. These
findings provide insights for modeling within-class variation in
speech-based AD detection.
\end{abstract}

\section{Introduction}

Neurocognitive disorders (NCD) such as Alzheimer's Disease (AD), present a substantial and growing challenge within the aging population, characterized by progressive cognitive decline across multiple domains, including memory, attention, and executive function \cite{lynch2020world}.
For timely intervention and management, in-person clinical assessments have been the primary protocol for screening AD patients in populations, where participants are examined using specially designed assessment tasks to test potential abnormal declines in cognitive abilities\cite{cookie1, cookie2, rabbit}.
In contrast to traditional on-set assessments, recent advancements in machine learning technologies have facilitated speech-based automatic AD detection as a promising screening approach, with the advantages of being scalable, accessible, and cost-efficient.

Common practices of machine learning AD detection are to model this task as binary classifications, i.e., prediction models are trained on audio recordings or transcripts from assessment tasks to classify participants as AD or non-AD \cite{ten_years, pipeline, kang2024not}.
This modeling largely inherits the paradigm of standard machine learning classification, which is dedicated to extracting discriminative features or patterns and then deploying a classifier for prediction.
In recent years, the progress in AD detection has been largely driven by the exploration of effective features and representation learning.
As a brief review, early works leveraged handcrafted acoustic and linguistic features.
For example, Alhanai et al.~\cite{2017audio} identified 12 acoustic features, including decreasing jitter, strongly associated with cognitive impairment. 
Winer and Frankenberg et al.~\cite{fraser2016linguistic, weiner2019speech} demonstrated the relevance of linguistic features such as parts-of-speech (POS) and word categories to AD.
More recently, the development of pre-trained models has mitigated the challenge in data-scarce tasks, leading to extensive research on deep embeddings for AD detection, encompassing speech-based \cite{koo2020exploiting, haulcy2021classifying, jinchao23, wavbert}, text-based \cite{balagopalan2020bert, yuan2020disfluencies, martinc2021temporal, yi22}, and multi-modal approaches \cite{koo2020exploiting, syed2021automated, titan}.


\begin{figure}[tbp]
\begin{center}
\hspace{-5pt}
\includegraphics[width=1\linewidth,scale=1.00]{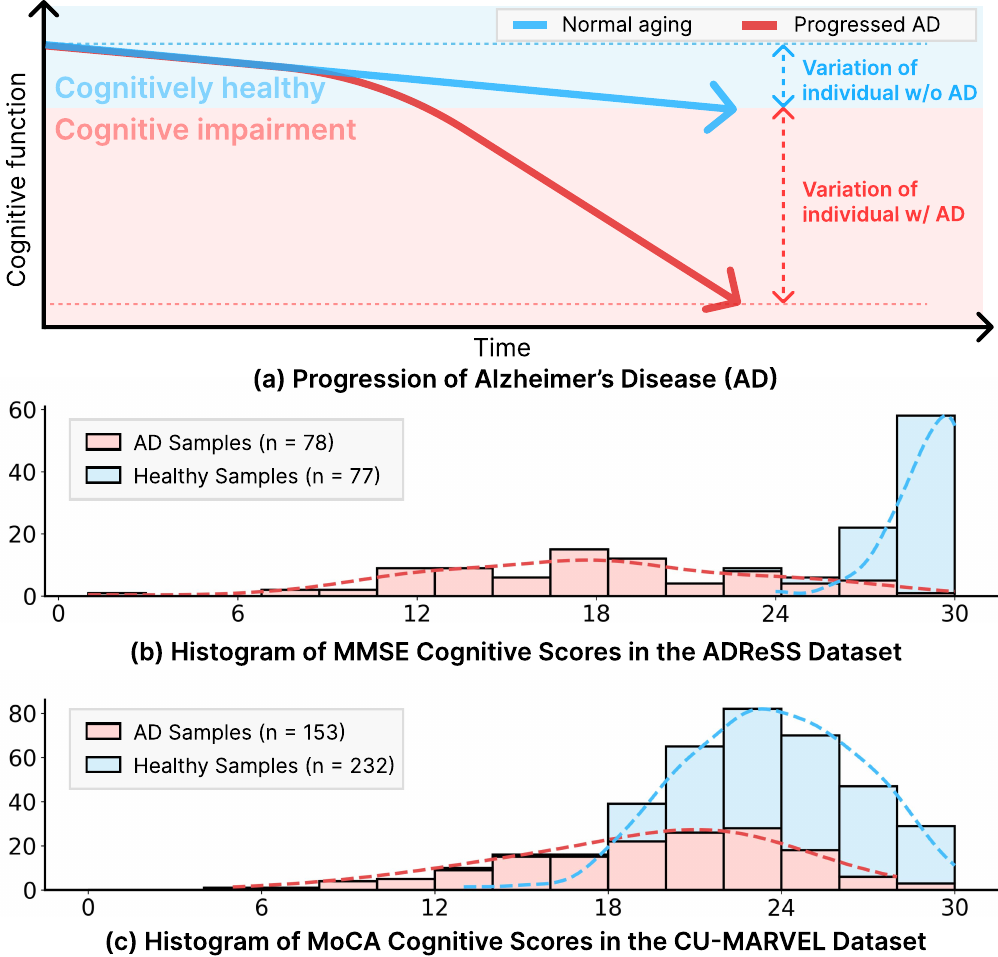}
\end{center}
\vspace{-15pt}
\caption{
Visualizations of the within-class variation in Alzheimer's disease (AD) detection.
The cognitive ability of AD individuals displays significant variation compared to healthy individuals \cite{ADlongitude1}.
}
\vspace{-19.15pt}
\label{fig:highlight}
\end{figure}

Beyond the challenge of feature extraction and representation learning, we propose that AD detection faces unique difficulties inherent to the nature of Alzheimer's Disease.
Medical literature \cite{ADlongitude1, ADlongitude2, ADdefinition} establishes AD as a \textit{degenerative disorder characterized by a continuum of pathophysiological changes, resulting in gradual cognitive and functional decline}.
This spectrum of cognitive performance among AD patients is illustrated in Fig.~\ref{fig:highlight}(a).
Consequently, under standard classification modeling, samples with the same label may exhibit varying degrees of cognitive impairment patterns, which could lead to considerable variability in recognition features.
This distinguishes AD detection from classification tasks such as image classification, which typically involve more consistent features under clearly defined categories.
An ideal solution would be to model AD detection by regression or multi-way classification to capture this continuum.
However, granular labels are rarely provided for most cognitive assessment tasks, especially for open-source datasets.
While reference continuous scores are occasionally available (such as Mini-Mental State Examination scores), these scores are obtained from different tests that are distinct from the assessment in the data and can hardly be regarded as gold standard for supervision.
Therefore, binary classification remains the prevailing method for Alzheimer's Disease (AD) detection.





\vspace{-1pt}
This work takes a first step towards the within-class variation (WCV) issue in AD detection.
We inspect sample variability beyond binary labels using the English ADReSS\cite{adress} and Cantonese CU-MARVEL \cite{pipeline} datasets (see Fig.~\ref{fig:highlight} (b) and Fig.~\ref{fig:highlight} (c)).
Even with comparable class members, AD samples exhibit a larger variance in cognitive function compared to healthy samples.
Given these findings, we propose that the conventional binary classification paradigm overlooks two critical aspects: a) \textit{within-class heterogeneity (WCH)}: samples with varying AD severity are assigned to the same class, potentially inhibiting the model's sensitivity to certain changes in cognitive function; and b) \textit{instance-level imbalance (ILI)}: the frequency of varying-severity samples is imbalanced even with balanced class size, thereby introducing potential bias.
A key challenge in tackling these issues is the lack of informative instance-level labels indicating sample severity.
Accordingly, this work explored sample score estimation that obtains sample-specific AD probabilities from ensemble models with only hard-label supervision. 
Building upon sample score estimation, we subsequently propose two approaches addressing WCH and ILI respectively: Soft Target Distillation (SoTD) and Instance-level Re-balancing (InRe).
The SoTD approach leverages the label refinery approach\cite{labelRefinery} to train classifiers using informative soft label supervision.
The InRe approach re-weights imbalanced samples at the instance level using log posterior ratios and inverse kernel density.

\vspace{-1pt}
We analyzed the proposed methods using the ADReSS and CU-MARVEL corpora.
The experimental results showed that the estimated sample scores aligned with the corresponding cognitive scores, despite the fact that cognitive scores were not available during training.
In addition, the InRe method guides model training to focus more on under-represented AD instances.
Finally, the proposed strategies exhibit remarkable performance improvements on both evaluation datasets.
This work presents an early investigation of the within-class variation issue in AD detection. 
We seek to provide helpful insights for developing more robust and reliable AD detection models.

\section{Problem Formulation}
\label{sec:formulation}
 
We consider AD detection under a latent progression view. Let $S$
denote a scalar coordinate of disease progression, $X$ the linguistic
observation, and $Y \in \{0,1\}$ the clinical diagnosis, with
$(x_i, y_i)$ denoting the observed pair for subject $i$. We model the
diagnostic probability along this coordinate as
\begin{equation}
r(s) = P(Y = 1 \mid S = s),
\label{eq:r}
\end{equation}
and assume that $r(s)$ is non-decreasing in $s$. Thus, subjects at
more advanced progression states are more likely to receive an AD
diagnosis. The binary label provides a coarse observation of the
latent progression state: subjects sharing the same diagnosis may
occupy different positions along $S$.
 
\subsection{Within-class heterogeneity}
\label{sec:wch}
 
Binary diagnosis preserves class membership while providing no
resolution among subjects within the same diagnostic class: the same
binary target is assigned despite potentially substantial differences
in progression state, as illustrated in Fig.~\ref{fig:highlight}. We express
this as an invariance property of the training objective.
 
Let $p_i$ denote the predicted AD probability for $x_i$, and we define
the corresponding AD log-odds as $a_i = \log \frac{p_i}{1-p_i}$.
Collecting these predictions in $\mathbf{a} = (a_1, \ldots, a_n)$, the
empirical binary cross-entropy risk is
\begin{equation}
\widehat{\mathcal{R}}(\mathbf{a}, \mathbf{y})
= \frac{1}{n} \sum_{i=1}^{n} \ell(a_i, y_i),
\label{eq:emprisk}
\end{equation}
where
$\ell(a_i, y_i) = -y_i \log p_i - (1 - y_i) \log (1 - p_i)$
and $p_i = \sigma(a_i)$.
Consider a label-preserving permutation $\pi$, such that
$y_{\pi(i)} = y_i$ for every $i$, with
$(\pi \mathbf{a})_i = a_{\pi(i)}$. Since all samples carrying the same
diagnosis share the same per-sample loss $\ell(\cdot, y)$, reassigning
a fixed set of predicted log-odds within a label class leaves the
empirical risk unchanged:
\begin{equation}
\widehat{\mathcal{R}}(\pi \mathbf{a}, \mathbf{y})
= \widehat{\mathcal{R}}(\mathbf{a}, \mathbf{y}).
\label{eq:perm}
\end{equation}
 
The hard-label objective therefore does not distinguish among
different arrangements of the predicted log-odds within the same
diagnostic class. It specifies whether a subject belongs to the AD or
non-AD class, but not how subjects within that class should differ
from one another.

\subsection{Instance-level imbalance}
\label{sec:ili}
 
Subjects within a diagnostic class occupy different positions along
$S$, and the objective does not specify how their predictions should
differ. The risk of a class aggregates these subjects according to how
the class populates the progression coordinate:
\begin{equation}
\mathcal{R}(\theta) = \sum_{y \in \{0,1\}} P(Y = y)
\int p(s \mid y) \,
\mathcal{R}(\theta \mid S = s, Y = y) \, ds ,
\label{eq:decomp}
\end{equation}
where $p(s \mid y)$ is the conditional density of progression states within
class $y$.
 
The class proportion $P(Y = y)$ determines the overall contribution of
each diagnostic class, whereas $p(s \mid y)$ determines how that
contribution is distributed across progression regions within the
class. Under this formulation, the imbalance within class enters the risk analogously to class-level imbalance: progression states that
are sparsely represented contribute little to the aggregate risk, and
errors concentrated on subjects in those states leave the objective
largely unchanged. Conventional class balancing operates on
$P(Y = y)$ and leaves $p(s \mid y)$ unchanged, so that a dataset can
be balanced between AD and non-AD while its AD subjects remain
unevenly distributed along the progression coordinate.

\section{Approaches}

This section presents the proposed approaches for addressing the two
aspects of within-class variation formulated above. Since the latent
progression coordinate $S$ is unavailable during training, we first
estimate sample-specific AD class probabilities as sample scores. We then develop Soft Target Distillation (SoTD) and
Instance-level Re-balancing (InRe) to address within-class
heterogeneity and instance-level imbalance, respectively.

\subsection{Sample score estimation}
\label{sec:score}
For binary cross-entropy, the population conditional risk is minimized
by the true class probability $P(Y=1\mid X=x)$
\cite{gneiting2007strictly}. A probabilistic classifier trained with
this objective can therefore provide a continuous estimate of the AD
class probability. 
We use this estimated probability as the sample-specific score for subsequent modeling of within-class variation.

\begin{figure}[t]
\begin{center}
\includegraphics[width=1\linewidth]{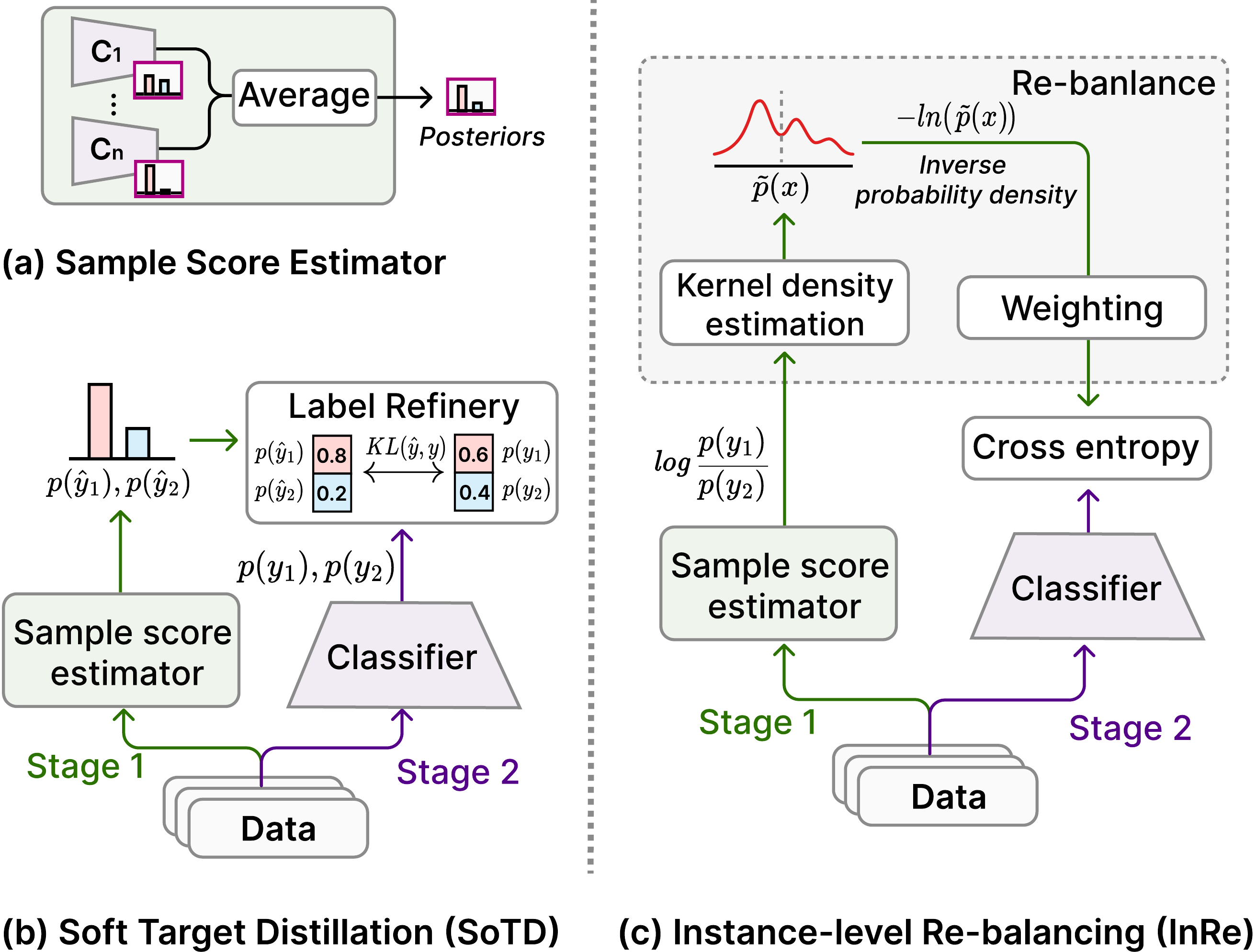}
\end{center}
\vspace{-15.5pt}
\caption{
Illustration of the proposed approaches.
KL(·) represents the KL divergence function.
$C_1$ to $C_n$ represent component classifiers.
}
\vspace{-17.5pt}
\label{fig:approach}
\end{figure}

In practice, we estimate the sample score using an ensemble of binary
classifiers. For subject $i$, let $q_i^{(m)}$ denote the estimated AD
class probability produced by the $m$-th component classifier. The
sample score is obtained by averaging the predictions of $M$ component
classifiers,
\begin{equation}
q_i = \frac{1}{M}\sum_{m=1}^{M} q_i^{(m)}.
\label{eq:sample_score}
\end{equation}
This ensemble estimation reduces variation associated with individual
component models and provides the sample-specific probabilities used
by the subsequent approaches, as illustrated in
Fig.~\ref{fig:approach}(a).

\subsection{Soft target distillation}
\label{sec:sotd}
To address within-class heterogeneity, Soft Target Distillation (SoTD)
replaces the class-shared binary target with the sample-specific
probability $q_i$ estimated above. Let $p_i$ denote the AD probability
predicted by the Stage-2 classifier. Following label refinery
\cite{labelRefinery}, we train the classifier using $q_i$ as
the soft target by minimizing
\begin{equation}
\mathcal{L}_{\mathrm{SoTD}}(p_i,q_i)
=
p_i \log \frac{p_i}{q_i}
+
(1-p_i)\log\frac{1-p_i}{1-q_i}.
\label{eq:sotd}
\end{equation}
For each sample, this divergence is uniquely minimized at $p_i=q_i$.
Equivalently, the corresponding AD log-odds satisfy
\begin{equation}
a_i^\star=\operatorname{logit}(q_i).
\label{eq:sotd_opt}
\end{equation}
Thus, same-class subjects with different sample scores are assigned
different target log-odds, thereby introducing sample-specific target resolution that is absent under class-shared hard-label supervision.

During Stage 2, the classifier is trained only with the sample-specific
soft targets and does not use the original hard labels. This follows
the multi-stage label-refinery setting \cite{labelRefinery},
which has also been applied to long-tailed recognition
\cite{li2021self}. An alternative is to combine the soft-target loss
with standard binary cross-entropy on the hard labels. However, our preliminary experiments and prior work \cite{labelRefinery}
suggest no additional benefit from this combination, and we therefore
use the soft targets alone.


\subsection{Instance-level re-balancing}
\label{sec:inre}
Section~\ref{sec:ili} shows that instance-level imbalance arises from
uneven representation along the latent progression coordinate $S$.
Following the principle of class re-balancing, which adjusts sample contributions through re-sampling or loss re-weighting, we apply an analogous strategy to within-class imbalance. Since $S$ is unavailable during training, InRe operationalizes this strategy in the classifier-derived score space. For subject $i$, we transform the estimated sample score $q_i$ into the AD log-odds
\begin{equation}
l_i = \log\frac{q_i}{1-q_i}.
\label{eq:inre_logodds}
\end{equation}
The resulting coordinate preserves the ordering of the estimated AD class probabilities and is used to characterize how training samples are distributed in the score space.

Let $\mathcal{L}=\{l_1,\ldots,l_n\}$ denote the log-odds values of the
training samples. We estimate the sample density over this score space
using Gaussian kernel density estimation and denote the resulting
density at $l$ by $\widetilde{p}(l)$. Samples located in lower-density
regions receive larger weights through
\begin{equation}
w_i=-\log \widetilde{p}(l_i).
\label{eq:inre_weight}
\end{equation}
The resulting weights are multiplied by the corresponding sample losses during training, increasing the contribution of sparsely represented samples. The bandwidth of the Gaussian kernel controls the smoothness of the estimated density and, consequently, the variation of the resulting sample weights.

\vspace{-5pt}
\section{Experimental setup}

\vspace{-3pt}
\subsection{Dataset}
\vspace{-3pt}
\noindent \textbf{ADReSS}
This is a frequently utilized dataset for AD detection, which is derived from the ADReSS challenge \cite{adress}.
It encompasses speech recordings and manual transcriptions obtained from 156 participants engaged in the Cookie Thief picture description task \cite{cookie1}.
The dataset is partitioned into a training set consisting of 108 samples and a test set with 48 samples, and there is an equal distribution of positive and negative cases within these subsets.

\noindent \textbf{CU-MARVEL}
This is a Cantonese corpus that was developed for the study of neurocognitive disorder diseases.
It is composed of speech recordings from a sequence of cognitive assessment tasks. 
In this particular work, we specifically employed the data from the Rabbit Story task \cite{rabbit} within this corpus.
Because this task is centered around the spontaneous speech of the participants. 
We designate this subset as the \textit{CUMV-R} dataset. The CUMV-R dataset contains manual speech transcriptions from 385 participants, among which 153 samples are positive and 232 samples are negative.

\vspace{-3pt}
\subsection{AD detection model}
\vspace{-3pt}
Pre-trained language models have demonstrated remarkable performance in AD detection \cite{li2021comparative, pipeline, wang2022exploring}.
In this study, BERT-family models \cite{bert, roberta} were employed to extract linguistic features from transcribed speech data.
Subsequently, multi-layer perception (MLP) classifiers and the Cross-entropy loss function were utilized.
For ADReSS corpus, we emulated the work \cite{li2021comparative} used bert-base-uncased model\footnote{
https://huggingface.co/google-bert/bert-base-uncased
} as the feature extractor.
The classifier contains 2 hidden layers with sizes of (32, 16).
It was optimized using Adam optimizer with a learning rate of $1e-3$, a batch size of 16, and the training was carried out for 20 epochs.
Regarding this, we followed the work in \cite{pipeline} and used a Chinese RoBERTa model\footnote{
https://huggingface.co/hfl/chinese-roberta-wwm-ext
} \cite{chinese-roberta} as the feature extractor.
The classifier had similar configurations, but the hidden layer sizes were set to (64, 16) and the learning rate was $1e-4$.

Along with the vanilla AD detection models, we implemented two baseline systems. 
Unlike SoTD, we utilized model ensembling, where the posteriors of the model outputs were averaged for decision fusion. 
Compared to the InRe method, we employed resampling-based class re-balancing, with the class sample rate set as the reciprocal of class frequencies.


\vspace{-3pt}
\subsection{Proposed approaches}
\vspace{-3pt}
We incorporated 5 component models in the sample score estimator and the model ensembling baseline.
As for density estimation in the InRe approach, the bandwidth was set to 2 by default.



\section{Results and discussions}
\subsection{Validate estimated sample scores}

\begin{figure}[tbp]
\begin{center}
\vspace{-8pt}
\includegraphics[width=0.95\linewidth,scale=1.00]{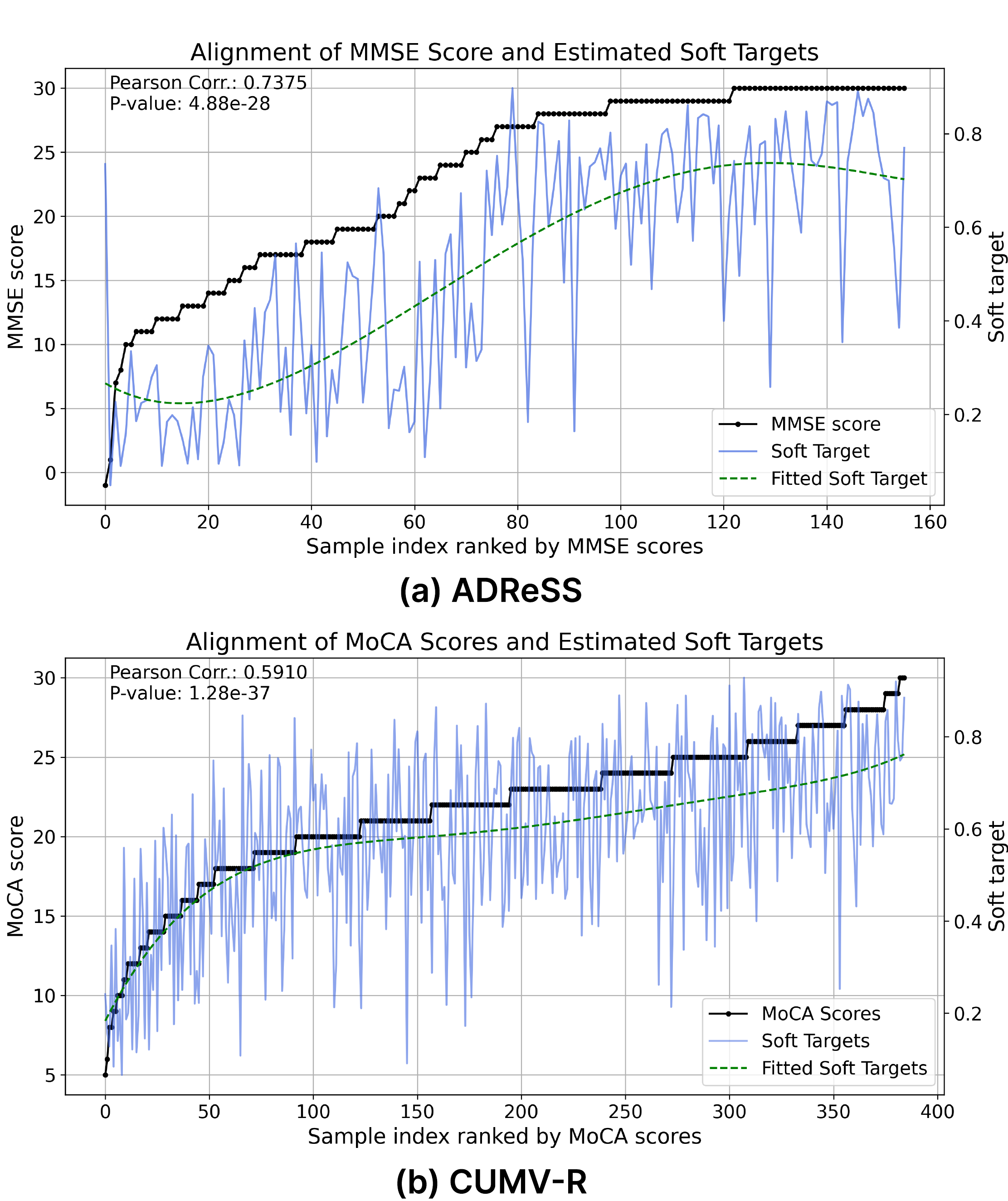}
\end{center}
\vspace{-20pt}
\caption{
The alignment between cognitive scores and estimated soft targets in the ADReSS (a) and CUMV-R (b) datasets.
}
\vspace{-12pt}
\label{fig:score}
\end{figure}

An effective sample score estimation serves as a prerequisite for
modeling within-class heterogeneity and imbalance using SoTD and InRe.
We therefore examine the estimated scores against participants'
cognitive assessment scores, i.e., MMSE scores in ADReSS and MoCA
scores in CUMV-R. These cognitive scores serve only as auxiliary
indicators of participants' cognitive ability and were not used during
model training.

As shown in Fig.~\ref{fig:score}, the estimated sample scores exhibit
clear alignment with the cognitive scores. Their Pearson correlation
coefficients are $0.7375$ and $0.5910$ on ADReSS and CUMV-R,
respectively. This empirical alignment suggests that the estimated AD
class probabilities capture information related to cognitive
variation beyond the binary diagnostic labels, supporting their use as
sample-specific scores in the proposed approaches. Although individual
scores show deviations from the corresponding cognitive assessments,
the overall correlations remain evident across both datasets.

\subsection{AD detection performance}
\begin{table}[tbp]
\centering
\caption{
 10-fold cross-validation results for AD detection using the ADReSS dataset. Each result is an average of 20 random runs. ``B-Acc.'' refers to balanced accuracy, and asterisk (*) indicates the proposed methods.
}
\vspace{-10pt}
\scalebox{0.99}{
\renewcommand{\arraystretch}{1.3} 
\hspace{-8pt}
\begin{tabular}{c|cccc}

\bottomrule
\multirow{2}{*}{} & \multicolumn{4}{c}{\textbf{ADReSS}} \\
\cline{2-5} 
&   \underline{B-Acc.} & \underline{F1} & Prec. & Recall  \\
\noalign{\hrule height 0.9pt}
Baseline 
& $.8330_{\pm .096}$  & $.8202_{ \pm .115}$ &.8371 &.8273 \\
\hline
Ensemble
& $.8399_{\pm .105}$	 & $.8246_{ \pm .113}$ &.8474	&.8271	 \\
SoTD*
&$\textbf{.8608}_{\pm .104}$	&	$\textbf{.8396}_{\pm .106}$ &\textbf{.8602} &	\textbf{.8382} \\
\hline
InRe*
&$\textbf{.8505}_{\pm .107}$ & $\textbf{.8351}_{ \pm .113} $ &.8286	&\textbf{.8573}	 \\
\bottomrule

\bottomrule
\multirow{2}{*}{} & \multicolumn{4}{c}{\textbf{CUMV-R}} \\
\cline{2-5} 
&  \underline{B-Acc.} & \underline{F1} & Prec. & Recall  \\
\noalign{\hrule height 0.9pt}
Baseline 
&$.6278_{\pm .067}$ & $.5072_{\pm .110}$ &.5993 &.4615 \\
\hline  
Ensemble
&$.6266_{	\pm .070}$ & $.5051_{ \pm .113}$ &.5983	&.4568	 \\
SoTD*
&$\textbf{.6356}_{\pm .065}$	&$\textbf{.5323}_{\pm .090}$ &.5893 &	\textbf{.4978} \\
\hline
Resampling
&$.6316_{\pm .074}$ & $.5330_{ \pm .142}$ &.5546	&.5640	 \\
InRe*
&$\textbf{.6407}_{\pm .074}$ & $\textbf{.5623}_{ \pm .099}$ &.5518	&\textbf{.5881}	 \\
\bottomrule

\end{tabular}
}

\label{tab:adress}
\vspace{-13pt}
\end{table}
Table \ref{tab:adress} presents our experimental results on ADReSS and CUMV-R corpus.
To reduce randomness, we performed 10-fold cross-validation for 20 random runs and reported the averaged results.
On the ADReSS corpus, the SoTD approach outperformed the baseline and ensemble systems across all 4 metrics.
This demonstrates general improvement in AD classification.
The InRE approach also led to improvement in the balanced accuracy and F1 metrics, while remarkably enhancing the recall rates.
This aligns with our expectation because the InRe approach could emphasize the infrequent positive (AD) samples in the dataset (as shown in Fig.~\ref{fig:highlight} (b)), thereby improving the model's sensitivity.
It is worth noting that \textit{this contribution differs from conventional class re-balancing methods since the number of positive and negative samples is equal in the training set.}
Thus, we contend that the advantage of the InRe approach stems its awareness of within-class imbalance.
The resampling approach was not carried out in this dataset, as this dataset is already balanced and the results will be the same as the baseline.

The CUMV-R dataset poses a more challenging scenario for several reasons: (a) it contains more variations in the cognitive scores; (b) positive and negative samples are more imbalanced (ratio=1.51); and (c) Cantonese BERT models might not be as well-developed as English models.
This is particularly evident in the low recall rates.
The SoTD approach outperform both baseline and ensembling methods by balancing the model's predictions, evidenced by the improved recall rate.
The resampling and InRe methods both show further improvement in recall rate, while the InRe method improved the models' sensitivity by a large margin ($0.4615\rightarrow0.5881$) and therefore remarkably improved the overall F1 score ($0.5072\rightarrow0.5623$).

\begin{figure}[ht]
\begin{center}
\vspace{-7pt}
\includegraphics[width=0.95\linewidth,scale=1.00]{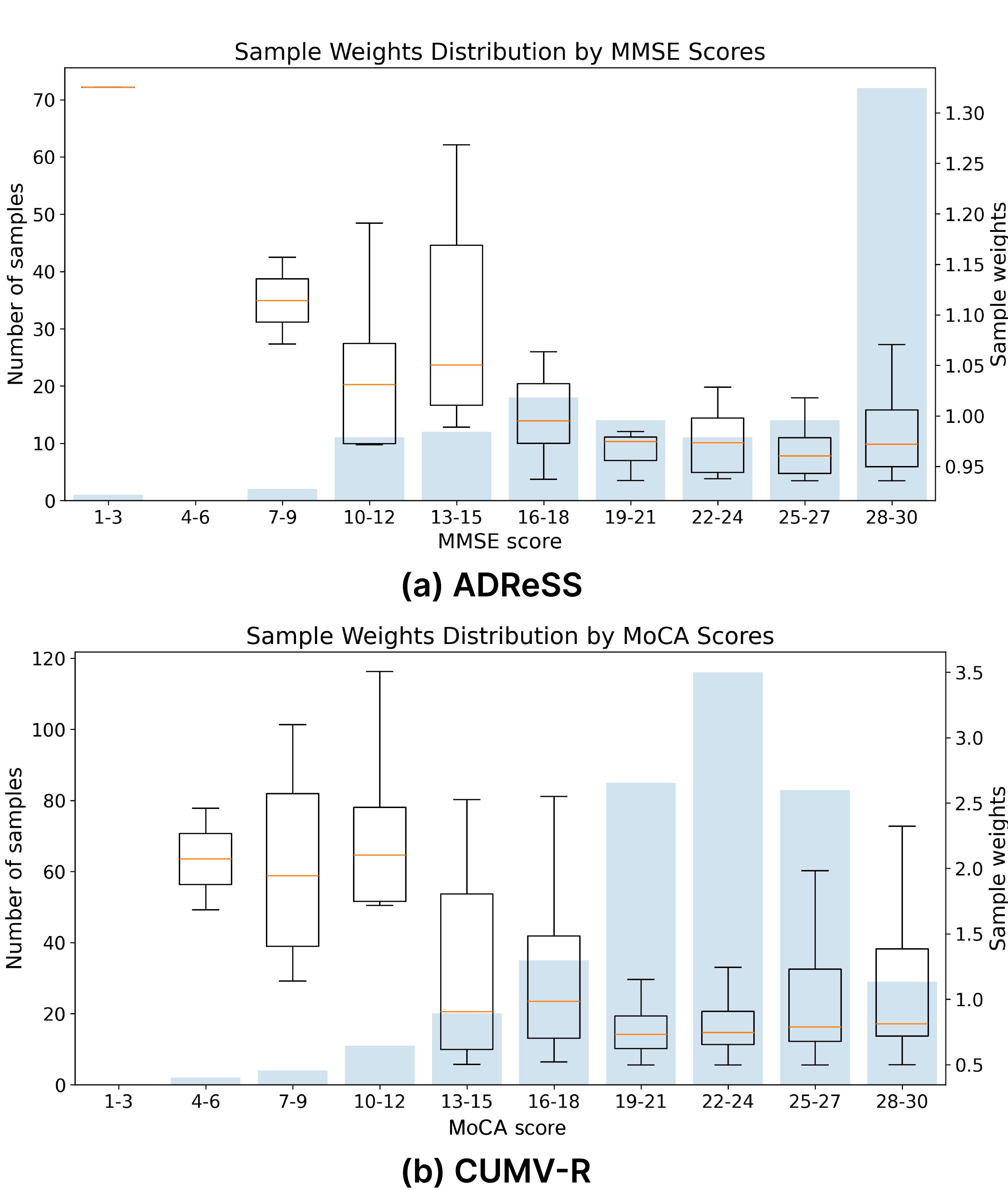}
\end{center}
\vspace{-20pt}
\caption{
Distribution of assigned sample weights across cognitive scores in the InRe method.
}
\vspace{-12pt}
\label{fig:weights}
\end{figure}

\subsection{Inspect sample weights in InRe approach}
To gain deeper insight into the InRe approach, a natural question is how sample weights are assigned across samples with varying cognitive levels.
In Fig.~\ref{fig:weights}, we collected all calculated sample weights and grouped them by corresponding cognitive score.
Again, cognitive scores are not accessible during the training phase. 
On the ADReSS corpus, the weights roughly correlate with the MMSE scores, but not that well.
We attribute this to the dispersion of sample groups in this dataset, where most groups contain fewer than 10 samples and can hardly guarantee statistical stability.
Nevertheless, samples with MMSE scores below 15 tend to be assigned notably higher weights.
In contrast, samples in the CUMV-R dataset display more distinct negative correlations -- the less frequent groups are assigned larger weights.
These findings imply that the InRe approach attempts to regulate the models' training to focus more on AD instances within sparse groups.
This behavior could explain the improvement in model sensitivity as presented in Table~\ref{tab:adress}.

\subsection{Effect of bandwidths in InRe}
Bandwidth is an important hyperparameter within the InRe approach during density estimation.
It controls estimation variance and bias, thereby affecting the sharpness of the sample weights distribution.
Table~\ref{tab:bins} presents a comparison of detection performance attained by employing different bandwidths. We found that using relatively \textit{smaller bandwidths} generally results in better performance, while the optimal result is obtained using a bandwidth of 2.
This could be because smaller bandwidths preserve more disparities among samples, while too small bandwidths (i.e., Bandwidth=1) incorporate potential noise in sample score estimation.

\begin{table}[htbp]
\vspace{-5pt}
\centering
\caption{
A comparison of different bandwidths adopted in the InRe approach on the CUMV-R dataset.
``B-Acc.'' refers to balanced accuracy.
}
\vspace{-10pt}
\scalebox{0.99}{
\renewcommand{\arraystretch}{1.3} 
\hspace{-8pt}
\begin{tabular}{c|cccccc}

\bottomrule
\multirow{2}{*}{} & \multicolumn{6}{c}{\textbf{Bandwidths}} \\
\cline{2-7} 
&   1 & 2 & 4 & 8 & 16 & 64 \\
\noalign{\hrule height 0.9pt}
B-Acc.
&.6357  &\textbf{.6407} &.6330 &.6367 &.6303 &.6241\\
\hline
F1
&.5582	&\textbf{.5623}  &.5528 &.5607	&.5524 &.5255	 \\
\hline
Precision
&.5459  &\textbf{.5518} &.5412 & .5450 & .5398 & .5508 \\
\hline
Recall
&.5867  &.5881 &.5804 & \textbf{.5925} & .5800 & .5169 \\
\bottomrule

\end{tabular}
}

\label{tab:bins}
\end{table}

\subsection{Sample logits distribution in SoTD \protect\footnote{This section was omitted from the conference version due to page constraints.} }

\begin{figure}[t]
\begin{center}
\includegraphics[width=0.95\linewidth,scale=1.00]{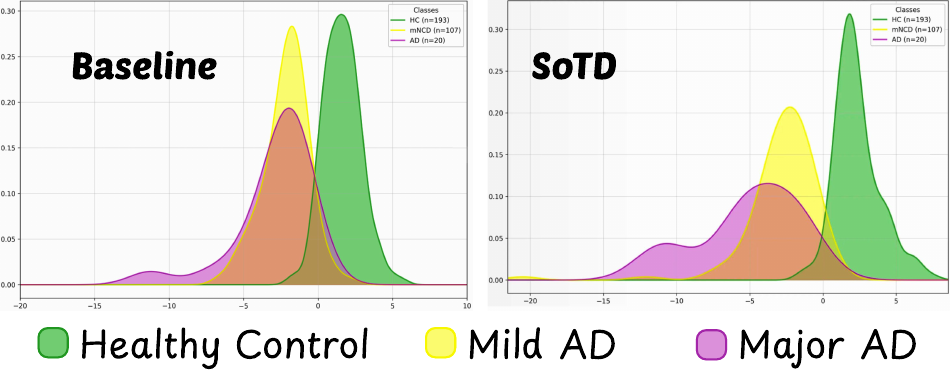}
\end{center}
\vspace{-15pt}
\caption{
Distribution of pre-softmax logits in the baseline and SoTD models, as a comparison of their discriminability of within-class sub-groups.
}
\vspace{-10pt}
\label{fig:dist}
\end{figure}

To further prob the inner working of the SoTD approach, Fig.~\ref{fig:dist} depicts the logits (pre-softmax values) distributions of the baseline and proposed SoTD models, allowing us to examine their learned
within-class representations.
Specifically, this experiment was conducted on the CUMV-R dataset, which provides additional 3-way labels: Healthy Control, Mild AD, and Major AD.
In both systems, the models are trained on \textit{binary classification} where we group "Mild AD" and "Major AD" as one "AD" group, in contrast to the "Healthy Control" group.
Therefore, Mild AD and Major AD form implicit within-class subgroups
during training.
During inference, the logits of all three groups are recorded, and their distribution are shown in Fig.~\ref{fig:dist}.

For the baseline model, the logit distributions of the Mild and Major
AD subgroups are almost fully overlapped. Under hard-label supervision,
these two subgroups share the same binary target, and the objective
provides no explicit supervision for distinguishing subjects within
the AD class, as discussed in Section~\ref{sec:wch}.
As a comparison, the subgroups can be relatively better distinguished by the SoTD model, as the purple and yellow distributions are less overlapped.
Despite the reduced overlap, substantial overlap remains between the two AD subgroups, suggesting that within-class variation remains a challenging problem for AD detection.

\section{Conclusions}
This work explores the issue of within-class variation in Alzheimer's Disease (AD) detection.
We posit that binary classification may overlook two crucial aspects: within-class heterogeneity (WCH) and instance-level imbalance (ILI).
We further introduce two simple yet effective methods to address these problems: soft target distillation (SoTD) and instance-level re-balancing (InRe).
Experiments on ADReSS and CU-MARVEL corpora demonstrated their advantages in detection performance.
Future work will explore the combination of the SoTD and InRe methods, and investigate the incorporation of cognitive scores during model training.


\vspace{-5pt}
\section{Acknowledgements}
This work is supported by the HKSARG Research Grants Council’s Theme-based Research Grant Scheme (Project No. T45-407/19N) and the CUHK Stanley Ho Big Data Decision Research Centre.

\newpage

\bibliographystyle{IEEEtran}
\bibliography{myRef}

@article{lynch2020world,
  title={World Alzheimer Report 2019: Attitudes to dementia, a global survey: Public health: Engaging people in ADRD research},
  author={Lynch, Chris},
  journal={Alzheimer's \& Dementia},
  year={2020},
  publisher={Wiley Online Library}
}

@article{cookie1,
  title={Performance on the Boston Cookie Theft picture description task in patients with early dementia of the Alzheimer's type: missing information},
  author={Giles, Elaine and Patterson, Karalyn and Hodges, John R},
  journal={Aphasiology},
  volume={10},
  number={4},
  pages={395--408},
  year={1996},
  publisher={Taylor \& Francis}
}

@book{cookie2,
  title={BDAE-3: Boston Diagnostic Aphasia Examination-Third Edition},
  author={Goodglass, Harold and Kaplan, Edith and Barresi, Barbara},
  year={2001},
  publisher={Lippincott Williams \& Wilkins Philadelphia}
}

@article{rabbit,
  title={``Frog, where are you?'' Narratives in children with specific language impairment, early focal brain injury, and Williams syndrome},
  author={Reilly, Judy and Losh, Molly and Bellugi, Ursula and Wulfeck, Beverly},
  journal={Brain and language},
  volume={88},
  number={2},
  pages={229--247},
  year={2004},
  publisher={Elsevier}
}

@article{ten_years,
  title={Ten years of research on automatic voice and speech analysis of people with Alzheimer's disease and mild cognitive impairment: a systematic review article},
  author={Mart{\'\i}nez-Nicol{\'a}s, Israel and Llorente, Thide E and Mart{\'\i}nez-S{\'a}nchez, Francisco and Meil{\'a}n, Juan Jos{\'e} G},
  journal={Frontiers in Psychology},
  volume={12},
  pages={620251},
  year={2021},
  publisher={Frontiers Media SA}
}

@misc{2017audio,
      title={Spoken Language Biomarkers for Detecting Cognitive Impairment}, 
      author={Tuka Alhanai and Rhoda Au and James Glass},
      year={2017},
      eprint={1710.07551},
      archivePrefix={arXiv},
      primaryClass={cs.AI}
}

@article{fraser2016linguistic,
  title={Linguistic features identify Alzheimer’s disease in narrative speech},
  author={Fraser, Kathleen C and Meltzer, Jed A and Rudzicz, Frank},
  journal={Journal of Alzheimer's Disease},
  year={2016},
  publisher={IOS Press}
}

@inproceedings{weiner2019speech,
  title={Speech reveals future risk of developing dementia: Predictive dementia screening from biographic interviews},
  author={Weiner, Jochen and Frankenberg, Claudia and Schr{\"o}der, Johannes and Schultz, Tanja},
  booktitle={ASRU},
  year={2019},
  organization={IEEE}
}

@inproceedings{koo2020exploiting,
  title={Exploiting Multi-Modal Features from Pre-Trained Networks for Alzheimer's Dementia Recognition},
  author={Koo, Junghyun and Lee, Jie Hwan and Pyo, Jaewoo and Jo, Yujin and Lee, Kyogu},
  booktitle={INTERSPEECH},
  year={2020}
}

@inproceedings{wavbert,
  title={Wavbert: Exploiting semantic and non-semantic speech using wav2vec and bert for dementia detection},
  author={Zhu, Youxiang and Obyat, Abdelrahman and Liang, Xiaohui and Batsis, John A and Roth, Robert M},
  booktitle={Interspeech},
  volume={2021},
  pages={3790},
  year={2021},
  organization={NIH Public Access}
}

@inproceedings{jinchao23,
  title={Leveraging pretrained representations with task-related keywords for alzheimer’s disease detection},
  author={Li, Jinchao and Song, Kaitao and Li, Junan and Zheng, Bo and Li, Dongsheng and Wu, Xixin and Liu, Xunying and Meng, Helen},
  booktitle={ICASSP 2023-2023 IEEE International Conference on Acoustics, Speech and Signal Processing (ICASSP)},
  pages={1--5},
  year={2023},
  organization={IEEE}
}

@article{yi22,
  title={Exploring linguistic feature and model combination for speech recognition based automatic ad detection},
  author={Wang, Yi and Wang, Tianzi and Ye, Zi and Meng, Lingwei and Hu, Shoukang and Wu, Xixin and Liu, Xunying and Meng, Helen},
  journal={arXiv preprint arXiv:2206.13758},
  year={2022}
}

@article{syed2021automated,
  title={Automated recognition of Alzheimer’s dementia using bag-of-deep-features and model ensembling},
  author={Syed, Zafi Sherhan and Syed, Muhammad Shehram Shah and Lech, Margaret and Pirogova, Elena},
  journal={IEEE Access},
  year={2021},
  publisher={IEEE}
}

@article{balagopalan2020bert,
  title={To BERT or not to BERT: comparing speech and language-based approaches for Alzheimer's disease detection},
  author={Balagopalan, Aparna and Eyre, Benjamin and Rudzicz, Frank and Novikova, Jekaterina},
  journal={arXiv preprint arXiv:2008.01551},
  year={2020}
}

@inproceedings{yuan2020disfluencies,
  title={Disfluencies and Fine-Tuning Pre-Trained Language Models for Detection of Alzheimer's Disease.},
  author={Yuan, Jiahong and Bian, Yuchen and Cai, Xingyu and Huang, Jiaji and Ye, Zheng and Church, Kenneth},
  booktitle={INTERSPEECH},
  year={2020}
}

@article{haulcy2021classifying,
  title={Classifying Alzheimer's disease using audio and text-based representations of speech},
  author={Haulcy, R'mani and Glass, James},
  journal={Frontiers in Psychology},
  volume={11},
  pages={624137},
  year={2021},
  publisher={Frontiers Media SA}
}

@article{martinc2021temporal,
  title={Temporal integration of text transcripts and acoustic features for Alzheimer's diagnosis based on spontaneous speech},
  author={Martinc, Matej and Haider, Fasih and Pollak, Senja and Luz, Saturnino},
  journal={Frontiers in Aging Neuroscience},
  year={2021},
  publisher={Frontiers}
}

@article{ADlongitude1,
  title={Longitudinal study of the transition from healthy aging to Alzheimer disease},
  author={Johnson, David K and Storandt, Martha and Morris, John C and Galvin, James E},
  journal={Archives of neurology},
  volume={66},
  number={10},
  pages={1254--1259},
  year={2009},
  publisher={American Medical Association}
}

@article{ADlongitude2,
  title={On the path to 2025: understanding the Alzheimer’s disease continuum},
  author={Aisen, Paul S and Cummings, Jeffrey and Jack, Clifford R and Morris, John C and Sperling, Reisa and Fr{\"o}lich, Lutz and Jones, Roy W and Dowsett, Sherie A and Matthews, Brandy R and Raskin, Joel and others},
  journal={Alzheimer's research \& therapy},
  volume={9},
  pages={1--10},
  year={2017},
  publisher={Springer}
}

@article{ADdefinition,
  title={Revising the definition of Alzheimer's disease: a new lexicon},
  author={Dubois, Bruno and Feldman, Howard H and Jacova, Claudia and Cummings, Jeffrey L and DeKosky, Steven T and Barberger-Gateau, Pascale and Delacourte, Andr{\'e} and Frisoni, Giovanni and Fox, Nick C and Galasko, Douglas and others},
  journal={The Lancet Neurology},
  volume={9},
  number={11},
  pages={1118--1127},
  year={2010},
  publisher={Elsevier}
}

@article{adress,
  title={{Alzheimer’s Dementia Recognition Through Spontaneous Speech: The ADReSS Challenge}},
  author={Luz, Saturnino and Haider, Fasih and de la Fuente, Sofia and Fromm, Davida and MacWhinney, Brian},
  journal={INTERSPEECH},
  year={2020}
}

@article{labelRefinery,
  title={Label refinery: Improving imagenet classification through label progression},
  author={Bagherinezhad, Hessam and Horton, Maxwell and Rastegari, Mohammad and Farhadi, Ali},
  journal={arXiv preprint arXiv:1805.02641},
  year={2018}
}

@inproceedings{li2021self,
  title={Self supervision to distillation for long-tailed visual recognition},
  author={Li, Tianhao and Wang, Limin and Wu, Gangshan},
  booktitle={Proceedings of the IEEE/CVF international conference on computer vision},
  pages={630--639},
  year={2021}
}

@article{pipeline,
  title={Integrated and enhanced pipeline system to support spoken language analytics for screening neurocognitive disorders},
  author={Meng, Helen and Mak, Brian and Mak, Man-Wai and Fung, Helene and Gong, Xianmin and Kwok, Timothy and Liu, Xunying and Mok, Vincent and Wong, Patrick and Woo, Jean and others},
  journal={Interspeech},
  year={2023}
}

@inproceedings{li2021comparative,
  title={A comparative study of acoustic and linguistic features classification for alzheimer's disease detection},
  author={Li, Jinchao and Yu, Jianwei and Ye, Zi and Wong, Simon and Mak, Manwai and Mak, Brian and Liu, Xunying and Meng, Helen},
  booktitle={ICASSP},
  pages={6423--6427},
  year={2021},
  organization={IEEE}
}

@article{wang2022exploring,
  title={Exploring linguistic feature and model combination for speech recognition based automatic ad detection},
  author={Wang, Yi and Wang, Tianzi and Ye, Zi and Meng, Lingwei and Hu, Shoukang and Wu, Xixin and Liu, Xunying and Meng, Helen},
  journal={arXiv preprint arXiv:2206.13758},
  year={2022}
}

@misc{bert,
      title={{BERT}: Pre-training of Deep Bidirectional Transformers for Language Understanding}, 
      author={Jacob Devlin and Ming-Wei Chang and Kenton Lee and Kristina Toutanova},
      year={2019},
      eprint={1810.04805},
      archivePrefix={arXiv},
      primaryClass={cs.CL}
}

@misc{roberta,
      title={RoBERTa: A Robustly Optimized BERT Pretraining Approach}, 
      author={Yinhan Liu and Myle Ott and Naman Goyal and Jingfei Du and Mandar Joshi and Danqi Chen and Omer Levy and Mike Lewis and Luke Zettlemoyer and Veselin Stoyanov},
      year={2019},
      eprint={1907.11692},
      archivePrefix={arXiv},
      primaryClass={cs.CL}
}

@inproceedings{chinese-roberta,
    title = "Revisiting Pre-Trained Models for {C}hinese Natural Language Processing",
    author = "Cui, Yiming  and
      Che, Wanxiang  and
      Liu, Ting  and
      Qin, Bing  and
      Wang, Shijin  and
      Hu, Guoping",
    booktitle = "Proceedings of the 2020 Conference on Empirical Methods in Natural Language Processing: Findings",
    month = nov,
    year = "2020",
    address = "Online",
    publisher = "Association for Computational Linguistics",
    url = "https://www.aclweb.org/anthology/2020.findings-emnlp.58",
    pages = "657--668",
}

@article{titan,
  title={Detecting Neurocognitive Disorders through Analyses of Topic Evolution and Cross-modal Consistency in Visual-Stimulated Narratives},
  author={Li, Jinchao and Wang, Yuejiao and Li, Junan and Kang, Jiawen and Zheng, Bo and Wong, Simon and Mak, Brian and Fung, Helene and Woo, Jean and Mak, Man-Wai and others},
  journal={arXiv preprint arXiv:2501.03727},
  year={2025}
}

@inproceedings{kang2024not,
  title={Not All Errors Are Equal: Investigation of Speech Recognition Errors in Alzheimer's Disease Detection},
  author={Kang, Jiawen and Li, Junan and Li, Jinchao and Wu, Xixin and Meng, Helen},
  booktitle={2024 IEEE 14th International Symposium on Chinese Spoken Language Processing (ISCSLP)},
  pages={254--258},
  year={2024},
  organization={IEEE}
}

@article{gneiting2007strictly,
  title={Strictly proper scoring rules, prediction, and estimation},
  author={Gneiting, Tilmann and Raftery, Adrian E},
  journal={Journal of the American statistical Association},
  volume={102},
  number={477},
  pages={359--378},
  year={2007},
  publisher={Taylor \& Francis}
}

\end{document}